\documentclass[aps,pra,twocolumn,showpacs]{revtex4}%
\usepackage{amsmath}
\usepackage{amsfonts}
\usepackage{amssymb}
\usepackage{graphicx}%
\providecommand{\U}[1]{\protect\rule{.1in}{.1in}}

\begin{document}
\title{Population Dynamics in Cold Gases Resulting from the Long-Range Dipole-Dipole Interaction}
\author{A. Mandilara$^{1,2}$, V. M. Akulin$^{1}$, and P. Pillet$^{1}$}
\affiliation{$^{1}$Laboratoire Aim\'{e} Cotton, CNRS, Campus d'Orsay, 91405, Orsay, France}
\affiliation{$^{2}$Quantum Information and Communication, \'{E}cole Polytechnique,
Universit\'{e} Libre de Bruxelles, CP~165, 1050 Brussels, Belgium}

\begin{abstract}
We consider the effect of the long range dipole-dipole interaction on the
excitation exchange dynamics of cold two-level atomic gase in the conditions
where the size of the atomic cloud is large as compared to the wavelength of
the dipole transition. We show that this interaction results in population
redistribution across the atomic cloud and in specific spectra of the
spontaneous photons emitted at different angles with respect to the direction
 of atomic polarization.

\end{abstract}

\maketitle

\section{Introduction}

Cold atomic gases give a possibility to address collective atomic quantum
states, where an elementary atomic excitation is coherently distributed among
a large number of individual atoms. Such systems have been considered in the
context of Quantum Information \cite{Brion}, coherence protection \cite{F},
individual photon manipulations \cite{Lukin} etc. For cold Rydberg gases that
have large dipole moments, one has to allow for the effects associated with
strong dipole-dipole \ interaction among atoms \cite{Gallagher}%
-\cite{Weidenmuller}. Considering the regime typical of current experimental
setting, one usually takes into account the regular static dipole-dipole
interaction $V\sim1/R^{3}$ among pairs of the atoms, which from the view point
of Quantum Electrodynamics is a result of exchange by a virtual strongly
off-resonant vacuum photon with a typical wavelength of the order of
interatomic distance $R$ \cite{AkulinKarlov}. This sort of interatomic
coupling can be attributed neither to the short range interaction, nor to the
long range ones, since the integral $\int_{0}^{\infty}V(R)R^{2}\mathrm{d}R$
corresponding to the average binary interaction logarithmically diverges at
both upper and lower limits. This circumstance results in a number of
interesting dynamic phenomena \cite{PhysicaD}, such as incomplete decay of a
single atom population at the limit of long times, which resembles the
Anderson localization effect. In the same time, for the description one cannot
make use of the continuous media model and the mean field approximation, which
are applicable only for the long range interactions.

Among experiments with Rydberg atoms, there are some \cite{Mourachko} that
have been done in the regime where the size of a cloud of cold Rydberg atoms
is of the order or larger than the wavelength of the resonant dipole active
transition. In such conditions, the role of the radiation trapping,
super-radiance \cite{Haroche}, and the dynamic long-range dipole-dipole
interactions ($\sim1/R$) is dominant, and the continuous media model becomes
applicable. In other words quantum states of the atomic ensemble and that of
the resonant radiation field get strongly entangled, whereas the eigenstates
of the compound atoms+field system correspond to atomic states dressed in the
resonant radiation field. Driven by the curiosity we will now consider the
atomic population dynamics in the simplest version of such a process.\ We make
use of the mean-field approximation, and concentrate on an exactly soluble
case of an infinite uniform continuous two-level media, ignoring the
contribution of the static dipole-dipole interaction not conforming the
requirements of the mean field model.

We consider a static continuous media corresponding to the gas of
$\mathcal{N}$ two-level particles with the transition frequency $\omega$ and
polarization $\mathbf{d}$ along $z$-direction that are uniformly scattered in
space at fixed positions. Since the static dipole-dipole interaction between
the atoms can be viewed as a process intermediated by virtual photon exchange,
the effective Hamiltonian that governs the evolution of the combined system in
the rotating wave approximation, is
\begin{equation}
\widehat{H}=\sum_{k}\hbar ck\widehat{a}_{\mathbf{k}}^{+}\widehat
{a}_{\mathbf{k}}^{-}+\frac{\hbar\omega}{2}\widehat{\sigma}_{\mathbf{k}}%
^{z}+\hbar v_{\mathbf{k}}\left(  \widehat{a}_{\mathbf{k}}^{+}\widehat{\sigma
}_{\mathbf{k}}^{-}+\widehat{\sigma}_{\mathbf{k}}^{+}\widehat{a}_{\mathbf{k}%
}^{-}\right)  ,
\end{equation}
where%

\begin{equation}
\widehat{\sigma}_{\mathbf{k}}^{i}=\sum_{j}\widehat{\sigma}_{j}^{i}%
\mathrm{e}^{\pm\mathrm{i}\left(  \mathbf{kr}_{j}\right)  },\text{ for}\quad
i=+,-,z,
\end{equation}
are the collective operators of the exciton $\mathbf{k}$, $v_{\mathbf{k}%
}=U\sqrt{k}\sin\alpha$ is the coupling and $\alpha$ is the angle between the
direction of the polarization $\mathbf{d}$ and the normal to \ the photon
polarization plane. The coupling strength $U\sqrt{k}=d\sqrt{4\pi\hbar ckn}$ is
a product of the photon vacuum field strength $\mathcal{E}=\sqrt{4\pi\hbar
ck/\mathcal{V}}$ \ in a volume $\mathcal{V=N}/n$ and the collective atomic
dipole moment $d\sqrt{\mathcal{N}}$, where $d$ is the atomic transition dipole
moment, $c$ is the speed of light, and $k=\left\vert \mathbf{k}\right\vert $.

We assume that at time $t=0$, all but one particles are in the ground state.
The consideration is equally applicable to the case where a group of particles
locating in a volume with a typical size $a$, small as compared to the
resonant transition wavelength $2\pi c/\omega$, is in a coherent superposition
of the individual excited states, such that the total number of excitations is
one. In order to take advantage of the uniform distribution, we take states%
\begin{equation}
\left\vert \mathbf{k}\right\rangle =\frac{1}{\sqrt{\mathcal{N}}}%
\widehat{\sigma}_{\mathbf{k}}^{+}\left\vert \mathbf{0}\right\rangle \label{a}%
\end{equation}
corresponding to a given wavevector $\mathbf{k}$ as the basis set in the
collective Hilbert space. Here $\left\vert \mathbf{0}\right\rangle $ denotes
the vacuum where all particles are in the ground state, and the condition that
the average distance $n^{-1/3}$ among the neighboring particles corresponding
to the density $n$ is much shorter than the transition wavelength $2\pi
c/\omega$ is implicit.

Each state, (\ref{a}), interacts only with the photon of the same wavevector,
and the amplitudes $\psi_{\mathbf{k}}$ and $\varphi_{\mathbf{k}}$ of the
particle+field compound state $\left\vert \mathbf{k}\right\rangle _{c}%
=\psi_{\mathbf{k}}\left\vert \mathbf{k,0}\right\rangle +\varphi_{\mathbf{k}%
}\left\vert \mathbf{0},\mathbf{k}\right\rangle $ (where the second quantum
number corresponds to photons) satisfy the Schr\"{o}dinger equation%
\begin{align}
\mathrm{i}\hbar\dot{{\psi}}_{\mathbf{k}}  &  =U\sqrt{k}\sin\left(
\alpha\right)  \varphi_{\mathbf{k}}+\frac{1}{\left(  2\pi\right)  ^{3}}%
\delta(t)\label{ba}\\
\mathrm{i}\hbar\dot{{\varphi}}_{\mathbf{k}}  &  =\hbar(ck-\omega
)\varphi_{\mathbf{k}}+U\sqrt{k}\sin\left(  \alpha\right)  \psi_{\mathbf{k}}
\label{bb}%
\end{align}
where Dirac delta function $\delta(t)$ stands for the initial condition
corresponding to the excitation location in the origin, that is at the point
$\mathbf{r}=0$.

The exact solution of (\ref{ba}) and (\ref{bb})
\begin{align}
\psi_{\mathbf{k}} =  &  -\mathrm{e}^{\mathrm{i}t\frac{\omega-kc}{2}}%
\frac{\mathrm{i}\cos\left[  \Omega t\right]  +\frac{(\omega-kc)\sin\left[
\Omega t\right]  }{2\Omega}}{\left(  2\pi\right)  ^{3}}\\
\varphi_{\mathbf{k}} =  &  -\frac{U\sqrt{k}\sin\left(  \alpha\right)
}{\left(  2\pi\right)  ^{3}}\mathrm{e}^{\mathrm{i}t\frac{\omega-kc}{2}}%
\frac{\sin\left[  \Omega t\right]  }{\Omega} \label{c}%
\end{align}
with the Rabi frequency $\Omega=\sqrt{\left(  \frac{\omega-kc}{2}\right)
^{2}+k\left(  \frac{U}{\hbar}\right)  ^{2}\sin^{2}\alpha}$ suggests
\begin{align}
\psi(\mathbf{r})  &  =\int k^{2}\mathrm{e}^{-\mathrm{i}kr\cos\Theta
+\mathrm{i}t\frac{\omega-kc}{2}}\frac{\mathrm{i}\cos\left[  \Omega t\right]
+\frac{(\omega-kc)\sin\left[  \Omega t\right]  }{2\Omega}}{-\left(
2\pi\right)  ^{3}}\mathrm{d}\gamma\label{da}\\
\varphi(\mathbf{r})  &  =\int\frac{k^{2}\mathrm{e}^{-\mathrm{i}kr\cos
\Theta+\mathrm{i}t\frac{\omega-kc}{2}}}{-8\pi^{3}}U\sqrt{k}\sin\left(
\alpha\right)  \frac{\sin\left[  \Omega t\right]  }{\Omega}\mathrm{d}%
\gamma\label{db}%
\end{align}
for the coordinate dependent amplitudes, with $\mathrm{d}\gamma=\mathrm{d}%
k\mathrm{d}\Gamma$, where $\mathrm{d}\Gamma$ denotes the integration over
solid angle and the integration over $k$ starts at the point $0$ and goes to
$+\infty$.

\begin{figure}[h]
\begin{center}
\includegraphics[
height=1.6994in,
width=2.4785in
]{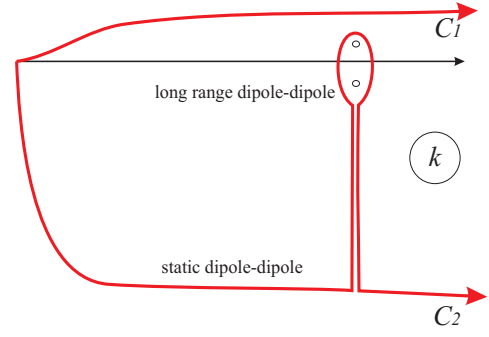}
\end{center}
\caption{Transformation of the integration contour for the inverse Fourier
transformation integrals (\ref{da})-(\ref{db}). The initial contour $C_{1}$
can be moved to the lower part of the complex plane of $k$. The part of the
contour $C_{2}$ circumventing the branching points (\ref{bp}) accounts for the
contribution of the long-range dipole-dipole interaction, while the remaining
part of the contour accounts for the regular static dipole-dipole interaction
$\sim1/R^{3}$.}%
\label{contour}%
\end{figure}

Long range dipole-dipole interaction corresponds to the exchange by photons
close to the resonance domain $k\simeq\omega/c$. In this domain, the
integrands (\ref{da})-(\ref{db}) have two branching points
\begin{equation}
\frac{c\omega\hbar^{2}-2U^{2}\sin^{2}\alpha\pm\mathrm{i}2U\sin\alpha
\sqrt{c\hbar^{2}\omega-U^{2}\sin^{2}\alpha}}{c^{2}\hbar^{2}} \label{bp}%
\end{equation}
in the complex plane of $k$ where $\Omega$ assumes zero value. Therefore the
contribution of the long range interaction is given by an integral along a
contour presented in figure~\ref{contour} circumventing these points, which can be expressed in terms of the
Bessel functions $J_{n}\left(  x\right)  $. In order to shorten the notations
we employ the units where $\omega/c=1$ henceforth. In the limit $U\ll1$ of a
coupling, which is weak as compared to the energy of atomic transition
quantum, the integration over $\mathrm{d}k$ yields%
\begin{align}
\psi(\mathbf{r})  &  =\int\frac{\mathrm{d}\Gamma}{2\pi^{2}}\frac{U_{\alpha
}\Theta_{H}\left(  r_{\Theta}\right)  \Theta_{H}(t-r_{\Theta})}{r_{\Theta
}^{1/2}(t-r_{\Theta})^{-1/2}}\times\label{ea}\\
&  J_{1}\left(  2U_{\alpha}\sqrt{r_{\Theta}(t-r_{\Theta})}\right)
\mathrm{e}^{-\mathrm{i}\left(  r_{\Theta}-tU_{\alpha}^{2}\right)  },\\
\varphi(\mathbf{r})  &  =\int\frac{\mathrm{d}\Gamma}{2\pi^{2}}\mathrm{i}%
U_{\alpha}\Theta_{H}\left(  r_{\Theta}\right)  \Theta_{H}(t-r_{\Theta}%
)\times\label{eb}\\
&  J_{0}\left(  2U_{\alpha}\sqrt{r_{\Theta}(t-r_{\Theta})}\right)
\mathrm{e}^{-\mathrm{i}\left(  r_{\Theta}-tU_{\alpha}^{2}\right)  },
\end{align}
up to terms of higher orders in $U$. Here $\Theta_{H}\left(  x\right)  $\ is
the Heaviside-$\Theta$ functions, while $r_{\Theta}$ and $U_{\alpha}$ stand
for $r\cos\Theta$ and $U\sin\alpha$, respectively.

\begin{figure}[h]
\begin{center}
\includegraphics[
height=2.5771in,
width=2.1197in
]{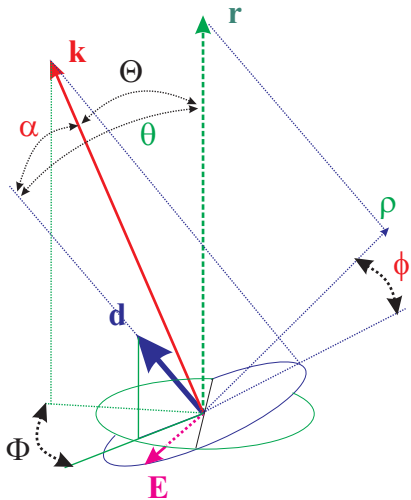}
\end{center}
\caption{Two sets of sperical coordinates in the space of the vavevectors
$\mathbf{k}$: the coordinates $\left\{  k,\alpha,\phi\right\}  $ associated
with the direction of polarization $\mathbf{d}$, and the coordinates $\left\{
k,\Theta,\Phi\right\}  $ associated with the radius vector $\mathbf{r}$. For a
given angle $\theta$ betweeen the direction of the polarization and the radius
vector, among these coordinates there exist relations (\ref{fa})-(\ref{fb}).
The photon polarization $\mathbf{E}$ is on the plane perpendicular to
$\mathbf{k}$.}%
\label{FIG.1}%
\end{figure}

Integration over the solid angle in (\ref{ea})-(\ref{eb}) has to be done
numerically in the polar coordinates associated either with the direction of
the polarization $\mathbf{d}$ (angles $\alpha$ and $\phi$) or with the
direction of the radius vector $\mathbf{r}$ (angles $\Theta$ and $\Phi$) as
shown in figure~\ref{FIG.1}. It has to be performed over the intervals
$\left\{  0,\pi\right\}  $ and $\left\{  0,2\pi\right\}  $, respectively.
Angles $\alpha$ and $\Theta$ do not belong to a same reference frame and
therefore one of these angles has to be expressed in the frame, to which
belongs the other one. For the purpose, one can make use of the relations%
\begin{align}
\sin^{2}\alpha &  =1-\left(  \cos\theta\cos\Theta+\sin\theta\sin\Theta\cos
\Phi\right)  ^{2}\label{fa}\\
\cos\Theta &  =\cos\theta\cos\alpha+\sin\theta\sin\alpha\cos\phi\label{fb}%
\end{align}
that follow from the invariance of the scalar product $\left(  \mathbf{kr}%
\right)  $. 

Alternatively one may employ the cylindrical set of coordinates $z=r\sin
\theta$, $\rho=r\cos\theta$, and $\phi$, associated with the direction of the
polarization $\mathbf{d}$.

\section{Population dynamics at the wavelength scale}

We note that there are two size scales in the problem: the wavelength $2\pi$
(that is $2\pi/k$ in the dimensional units) and a distance $1/U$ (that is
$c/U$ in the dimensional units) at which the light propagates during the Rabi
period at resonance. Population distribution at these scales have to be
analyzed separately.

\begin{figure}[h]
\begin{center}
\includegraphics[
height=1.9207in,
width=2.5002in
]{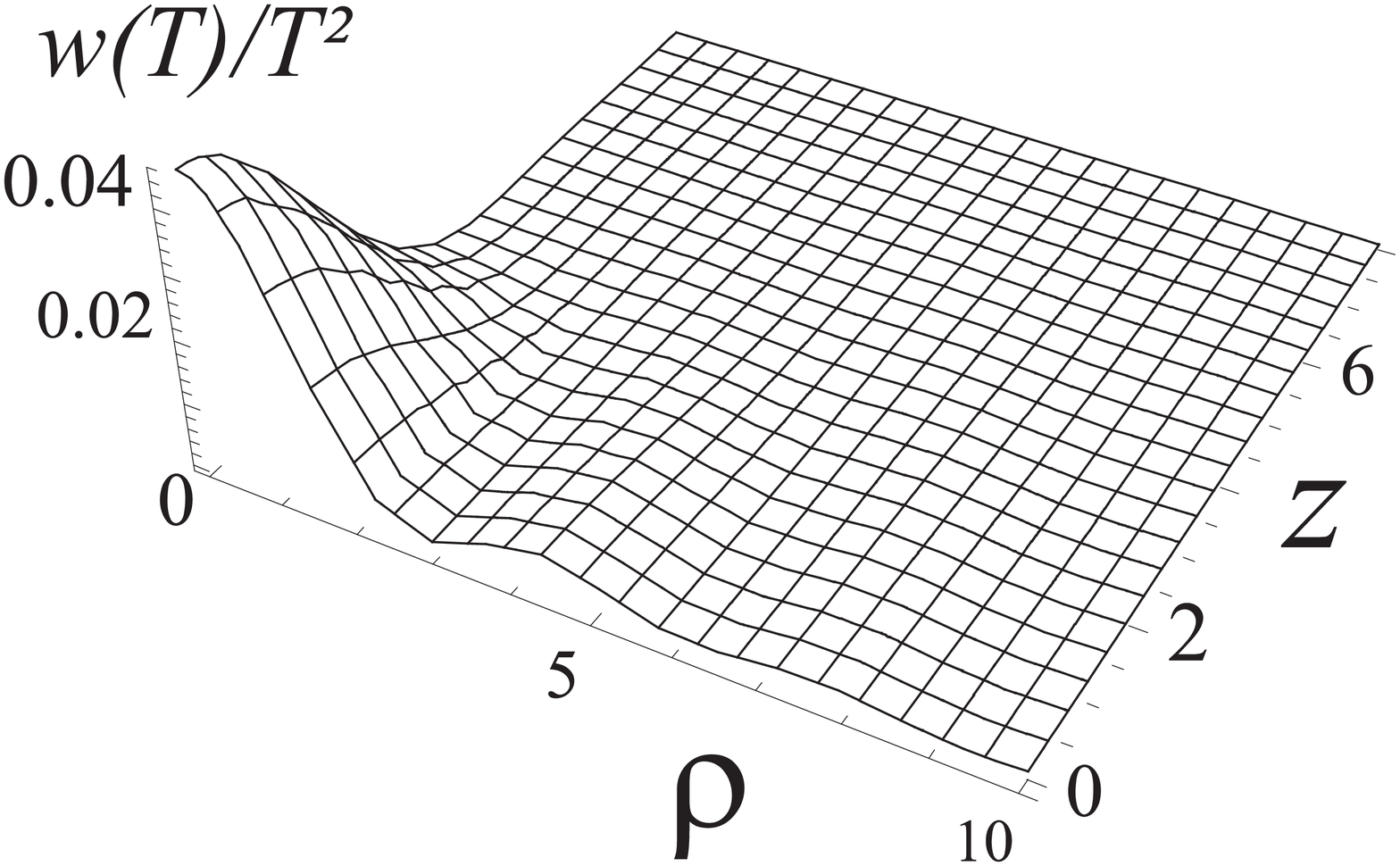}
\end{center}
\caption{Probability of the excitation $w(r)=\left\vert \psi(r)\right\vert
^{2}$ for the infinite media and for times $T=tU^{2}\ll1$ given by
Eq.(\ref{h}). }%
\label{Fig2}%
\end{figure}

Let us first consider the wavelength scale. For short times $t\ll1/U^{2}$, the
Bessel function can be cast in Taylor series and the first order yields
\begin{equation}
\psi(\mathbf{r})=\frac{tU^{2}}{2\pi^{2}}\int\sin^{2}\left(  \alpha\right)
\Theta_{H}\left(  \cos\Theta\right)  \mathrm{e}^{\mathrm{i}r\cos\Theta
}\mathrm{d}\Gamma\label{g}%
\end{equation}
for the amplitude in (\ref{eb}). With the help of equation (\ref{fa}) one can
calculate the integral exactly and find%
\begin{align}
\psi(\mathbf{r}) &  =\mathrm{i}\frac{tU^{2}}{4\pi}\frac{\left(  3r^{2}%
-2e^{-\mathrm{i}r}\left(  r^{2}+\mathrm{i}r+1\right)  +2\right)  }{r^{3}%
}+\nonumber\\
&  \mathrm{i}\frac{tU^{2}}{4\pi}\frac{\left(  r^{2}+2e^{-\mathrm{i}r}\left(
r^{2}-3\mathrm{i}r-3\right)  +6\right)  \cos(2\theta)}{r^{3}}\label{h}%
\end{align}
$\allowbreak$which results in the population $\left\vert \psi(\mathbf{r}%
)\right\vert ^{2}$ quadratically increasing in time, shown in
figure~\ref{Fig2}. Note, that at long distances the probability corresponding to
(\ref{h}) decreases as $r^{-2}$, which means that the integral of the
population over the volume logarithmically diverges at the upper limit being
truncated just at the radius $r=t$. In other words, some part of the
excitation is getting transferred by the resonant photons at a large distance
$r\sim t$. \ 

The distribution of figure~\ref{Fig2} persists as long as $tU^{2}$ remains a
small number. Its structure can be easily understood when one notice that
according to the second equation (\ref{c}), the amplitudes of emitted photons
are located in the domain of resonance where the the detunings $(1-k) $ are
small. These photons, after being emitted, result in discarding of the
corresponding spacial harmonics from the initial $\delta$-like excitation
profile $\psi(\mathbf{r})$ containing all harmonics with equal amplitudes. But
such a discarding implies establishing of the population distribution given by
the discarded harmonics taken with the opposite signs. As the result, the sum
of the $\delta$-like distribution and\ the latter distributions of
$\psi(\mathbf{r})$ no longer contains harmonics strongly coupled to the
cooperative resonant radiation. In course the of time the harmonics with
higher detuning are getting involved in the process, and the net population of
figure~\ref{Fig2} increases. For an initial distribution of a small but finite
size $a$, the relative amount of this population is given by the ratio of the
length $U^{2}t$ of the $k$-interval of emitted photons and the length
$\sim1/a$ of the $k$-interval occupied by the initial distribution.

\begin{figure}[h]
\begin{center}
\includegraphics[
height=3.1419in,
width=2.2866in]{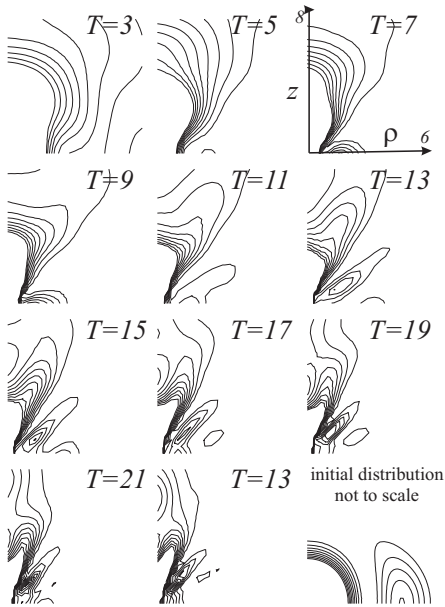}
\end{center}
\caption{Contour plot of the population distribution for long times $T=tU^{2}%
$.}%
\label{Fig3a}%
\end{figure}

In course of time $t \geq
U^{-2}$, the distribution $\left\vert \psi(\mathbf{r})\right\vert ^{2}$ in the
domain $r\sim1$ gets modified, as shown in figure~\ref{Fig3a}.

\begin{figure}[h]
\begin{center}
\includegraphics[height=1.721in,width=2.808in]{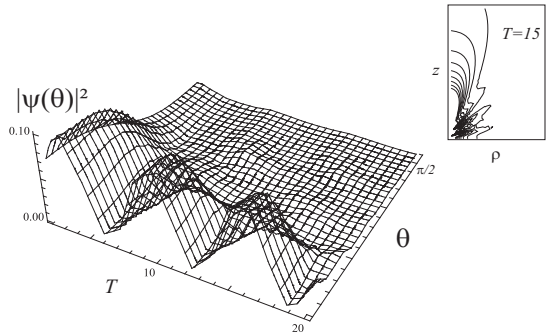}
\end{center}
\caption{Assymptotic dependence (\ref{k}) of the angular part of the
population distribution on time $T=tU^{2}$. In the inlet we show the contour
plot of the population distribution calculated with the assymptotic formula
for $T=15$.}%
\label{Fig5}%
\end{figure}

One can find an analytic expression for the asymptotic form of $\psi
(\mathbf{r})$ at $t\rightarrow\infty$. The main contribution to the integral
(\ref{eb}) comes from the domain of small $\cos\Theta$, that results from the
exchange by the resonant photons with $\mathbf{k}$ almost orthogonal to
$\mathbf{r}$. Integration over $\mathrm{d}\Gamma$ can be carried out in the
reference system associated with $\mathbf{r}$ with the allowance for the
approximate relation $\sin^{2}\alpha\simeq1-\left(  \sin\theta\cos\Phi\right)
^{2}$ and yields%
\begin{align}
\psi(\mathbf{r}) &  =\frac{\mathrm{i}}{2\pi\sqrt{z^{2}+\rho^{2}}}%
\mathrm{e}^{-\mathrm{i}\frac{tU^{2}}{2}\frac{2z^{2}+\rho^{2}}{z^{2}+\rho^{2}}%
}J_{0}\left(  \frac{tU^{2}\rho^{2}}{2\left(  z^{2}+\rho^{2}\right)  }\right)
\nonumber\\
&  -\frac{\mathrm{i}}{2\pi\sqrt{z^{2}+\rho^{2}}}\mathrm{e}^{-\mathrm{i}%
tU^{2}\frac{2z^{2}+\rho^{2}}{z^{2}+\rho^{2}}}J_{0}\left(  \frac{tU^{2}\rho
^{2}}{z^{2}+\rho^{2}}\right)  .\label{i}%
\end{align}
One sees that the population distribution decreases with the radius as
$1/r^{2}$ and has an angular dependence $\left\vert \psi(\theta)\right\vert
^{2}$
\begin{equation}
\left\vert \mathrm{e}^{\mathrm{i}\frac{tU^{2}}{2}\left(  2-\sin^{2}%
\theta\right)  }J_{0}\left(  \frac{tU^{2}\sin^{2}\theta}{2}\right)
-J_{0}\left(  tU^{2}\sin^{2}\theta\right)  \right\vert ^{2},\label{k}%
\end{equation}
shown in figure~\ref{Fig5}. Angular shrinking of the distribution in the course of time can be
understood if we note that the photons emitted in the direction close to that
of the particles polarization are weekly coupled due to the smallness of the
projection of the electric strength vector to the particle dipole moment.
Therefore the typical evolution time for their interaction with the atomic
cloud is longer. \begin{figure}[h]
\begin{center}
\includegraphics[height=2.6783in,width=2.5512in]{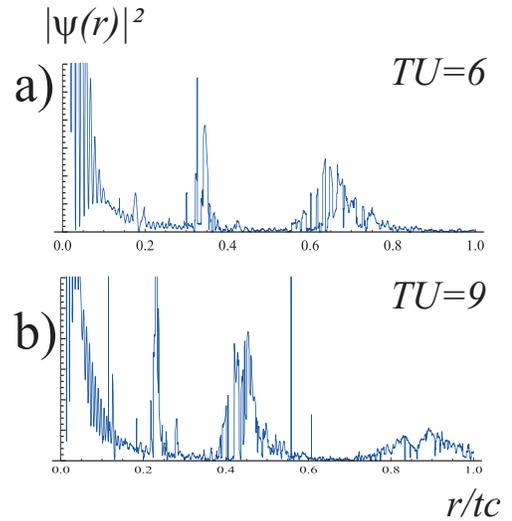}
\end{center}
\caption{Population distribution along the radial direction at distances
$\rho\sim c/U$ for two values of the parameter $TU=6$ (a) and $TU=9$ (b)
resulting from the numerical integration of equation (\ref{eb}).}%
\label{Fig3}%
\end{figure}

\section{Population dynamics and emitted radiation at large distance}

What happens at the large scale with typical distances $r\sim1/U$? The results
of numerical integration of (\ref{eb}) are shown in figure~\ref{Fig3} for two
different times. One sees rather irregular structures to some extend
resembling the time dependence of fractional revivals \cite{Averbukh}, which
results from the interference of different spacial harmonics with the
wavevectors close to unity. One should note that this regime implies that the
size of the system is larger or comparable with the distance $r\sim1/U$.
However, this distance can be relatively large, exceeding the typical size of
the atomic cloud. \begin{figure}[h]
\begin{center}
\includegraphics[height=1.7132in,width=2.2105in]{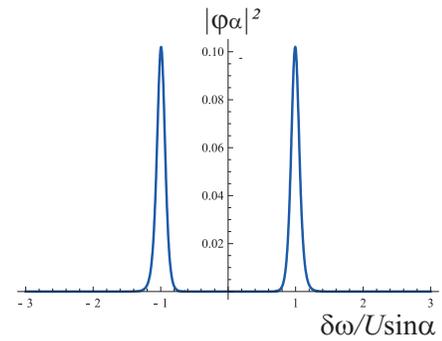}
\end{center}
\caption{Probability of photon detection (\ref{l}) for an atomic cloud of a
finite size $L=4c/\omega$ and for the atomic density profile $n(\mathbf{r}%
)\sim1/\cosh^{2}(\left\vert \mathbf{r}\right\vert /L)$.}%
\label{rad}%
\end{figure}In such a situation, instead of considering the population
distribution at distances $r\geq
1/U$, one may calculate the frequency spectrum of photons detected at a
distance much larger as compared to the typical size of the atomic cloud.

The calculations relevant to the case of large distances rely on the Fourier
transformed (\ref{ba}) and (\ref{bb})
\begin{align}
\varepsilon\psi_{\mathbf{k}}  &  =U\sqrt{k}\sin\left(  \alpha\right)
\varphi_{\mathbf{k}}+\frac{1}{\left(  2\pi\right)  ^{3}}\label{m}\\
\varepsilon\varphi_{\mathbf{k}}  &  =(k-1)\varphi_{\mathbf{k}}+U\sqrt{k}%
\sin\left(  \alpha\right)  \psi_{\mathbf{k}}, \label{n}%
\end{align}
while the resulting amplitude%
\begin{equation}
\varphi_{\mathbf{k}}=\frac{-\sqrt{k}U\sin\alpha}{8\pi^{3}\left(
-\varepsilon^{2}+k\varepsilon-\varepsilon+kU^{2}\sin^{2}\alpha\right)  }
\label{o}%
\end{equation}
to have a photon with the wavevector $\mathbf{k}$ at an angle $\alpha$ to the
direction of the polarization has to be averaged over an interval around $k=1$
of a width $\sim2\pi/L$ corresponding to the uncertainty resulting from the
finite size $L$ of the system. The averaging has to be performed with a weight
function $n(\kappa)$ given by the Fourier transformed particle density profile%
\begin{equation}
\varphi_{\alpha}=\int\frac{-n(\kappa)U\sin\alpha\ \ \mathrm{d}\kappa}{8\pi
^{3}\left(  -\varepsilon^{2}+\kappa\varepsilon+U^{2}\sin^{2}\alpha\right)  }.
\label{p}%
\end{equation}
For an analytic profile $n(\kappa)$ this yields the spectral intensity%
\begin{equation}
\left\vert \varphi_{\alpha}\right\vert ^{2}=\frac{1\ \ }{16\pi^{4}\left(
\delta\omega\right)  ^{2}}n^{2}\left(  \delta\omega-\frac{1}{\delta\omega
}\right)  , \label{l}%
\end{equation}
where the frequency deviation from the resonance $\delta\omega=\varepsilon
/U\sin\alpha$ is scaled by the interaction.

In figure~\ref{rad} we present the probability of the photon detection as a
function of frequency calculated for the particle density distribution
$n(\mathbf{r})=1/\cosh^{2}(\left\vert \mathbf{r}\right\vert /L)$ with $L=4$.
This distribution stands as a physical approximation to the delta function
spatial distribution of excited atoms we have assumed throughout the
derivation. One sees, that though the cooperative coupling between the
particles and the electromagnetic field does not govern the population
distribution $\left\vert \psi(\mathbf{r})\right\vert ^{2}$ over the ensemble
of two-level particles of a finite size $L<c/U$, it manifests itself in the
spectrum of emitted photons that can be registered at distances large as
compared to $L$.

\begin{figure}[h]
\begin{center}
\includegraphics[height=1.9199in,width=2.6628in]{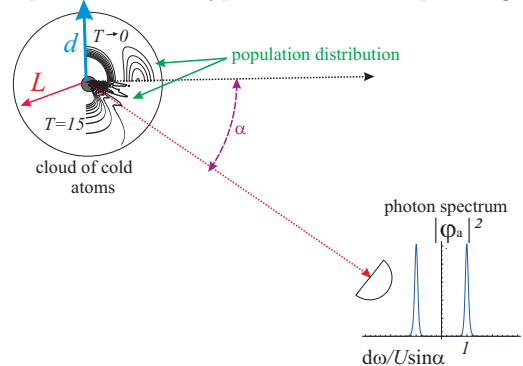}
\end{center}
\caption{The long range dipole-dipole interaction results in the
redistribution of the probability amplitude to be in the excited state, which
occurs with the typical rate $U^{2}$. The population distribution for small
times and for long times are shown by the contour plots in the upper and the
lower parts inside the circle representing the atomic cloud. In 3D the
distribution has cylindrical symmetry around the polarization axis
$\mathbf{d}$. The process is associated with the cooperative emission of
photons, and the typical scale of the spectrum of these photons is of the
order of $U$.}%
\label{overall}%
\end{figure}

\section{Conclusions}

We conclude by presenting the overall picture of the process shown in
figure~\ref{overall}. An atomic cloud, initially in the ground state, has been
excited in a small domain close to the center. In course of time the
excitations gets redistributed over all the volume of the cloud due to the
long range dipole-dipole interaction. The typical rate of the process
corresponds to the $U^{2}\rightarrow4\pi d^{2}n/\hbar$. For short times, the
population distribution is given by (\ref{h}) and for $tU^{2}\gg1$ it takes
the asymptotic form (\ref{i}). The process is associated with the cooperative
spontaneous emission of photons. The typical scale corresponding to the
spectrum of emitted photons is given by the interaction strength $U\rightarrow
d\sqrt{4\pi\omega n/\hbar}$, and at a large distance from the atomic cloud it
has a two-peak structure of equation (\ref{l}). As a general conclusion, one
may state that the presence of two essentially different energy scales ($U$
and $U^{2}$) can be considered as typical manifestation of collective atomic phenomena.

The authors are grateful to T. F.\ Gallagher for stimulating discussions.

\end{document}